\documentclass[aps,pra,twocolumn,superscriptaddress,amsmath,showpacs,amssymb]{revtex4}
\usepackage{bm}
\usepackage{amsmath,amssymb}
\usepackage[dvips,final]{graphicx}

\newcommand{\be}{\begin{equation}}
\newcommand{\ee}{\end{equation}}
\newcommand{\BM}{\begin{pmatrix}}
\newcommand{\EM}{\end{pmatrix}}

\bmdefine{\bx}{x}
\bmdefine{\by}{y}
\bmdefine{\bz}{z}
\bmdefine{\bmb}{b}
\renewcommand{\phi}{\varphi}
\newcommand{\MH}{\mathrm{H}}
\newcommand{\at}{\Tilde{a}}
\newcommand{\ab}{\Bar{a}}
\newcommand{\bt}{\Tilde{b}}
\newcommand{\bb}{\Bar{b}}
\newcommand{\T}[1]{\Tilde{#1}}
\newcommand{\xit}{\Tilde{\xi}}
\newcommand{\xib}{\Bar{\xi}}
\renewcommand{\d}{\dagger}
\renewcommand{\L}{\mathcal{L}}
\newcommand{\M}{\mathcal{M}}
\newcommand{\ve}{\varepsilon}
\newcommand{\bra}{\langle 0 |}
\newcommand{\ket}{| 0\rangle}
\newcommand{\tf}[6]{\begin{Bmatrix}#1\\#2\end{Bmatrix}^{#5}\begin{Bmatrix}#3\\#4\end{Bmatrix}^{#6}}
\newcommand{\Real}{\mathrm{Re}}
\begin{document}

\title{Derivation of non-Markoffian transport equations for trapped 
cold atoms in nonequilibrium thermal field theory} 
\date{\today}
\author{Y.~Nakamura}
\email{nakamura@aoni.waseda.jp}
\affiliation{Department of Electronic and Photonic Systems, Waseda
University, Tokyo 169-8555, Japan} 
\author{T.~Sunaga}
\email{tomoka@fuji.waseda.jp}
\affiliation{Department of Physics, Waseda University, Tokyo 169-8555,
Japan} 
\author{M.~Mine}
\email{mine@waseda.jp}
\affiliation{Waseda University Honjo Senior High School, 1136
Nishitomida, Honjo, Saitama 367-0035, Japan} 
\author{M.~Okumura}
\email{okumura.masahiko@jaea.go.jp}
\affiliation{CCSE, Japan Atomic Energy Agency, 6-9-3 Higashi-Ueno,
Taito-ku, Tokyo 110-0015, Japan} 
\affiliation{CREST(JST), 4-1-8 Honcho, Kawaguchi-shi, Saitama 332-0012,
Japan}  
\author{Y.~Yamanaka}
\email{yamanaka@waseda.jp}
\affiliation{Department of Electronic and Photonic Systems, Waseda
University, Tokyo 169-8555, Japan} 

\begin{abstract}
The non-Markoffian transport equations 
for the systems of cold Bose atoms 
confined by a external potential 
both without and with a 
Bose-Einstein condensate are derived in the framework of nonequilibrium 
thermal filed theory (Thermo Field Dynamics). 
Our key elements are an explicit particle representation and a 
self-consistent renormalization condition which are essential in thermal field theory.  
The non-Markoffian transport equation  for the non-condensed system, derived at the two-loop level, 
is reduced in the Markoffian limit to the ordinary quantum Boltzmann equation 
derived in the other methods.
For the condensed system, we derive a new transport equation 
with an additional collision term which becomes important 
in the Landau instability.
\end{abstract}
\pacs{03.75.Kk, 05.20.Dd, 05.30.Jp}
\maketitle
\section{Introduction}

The systems of trapped cold atoms are ideal for studying quantum
many-body theories such as quantum field theory and thermal field
theory. They are dilute and weak-interacting, so theoretical
calculations can be compared with experimental results
directly. 

Since the realization of Bose--Einstein condensates \cite{Cornell,
Ketterle, Bradley}, the formation and grow of condensate
\cite{Miesner}, the thermal shift of the energy spectrum \cite{Jin}, and
many other intriguing phenomena have been observed with good accuracy,
and offer opportunities to test quantum many-body theories in both
equilibrium and nonequilibrium. 

In the aim of describing the kinetics of the trapped cold atom system, a
number of theoretical approaches have been proposed such as the methods
of the quantum Boltzmann master equation \cite{Gardiner1, Gardiner2},
the quantum Boltzmann equation with the local density approximation
\cite{Luiten, Yamashita}, the closed path (CTP) formalism \cite{Stoof,
Zaremba}, and the effective Hamiltonian method in Thermo Field Dynamics
(TFD) \cite{Matsumoto}. These are in good agreement with the experiments
\cite{Ketterle, Miesner}. They however are based on a phase-space
distribution function, and the energy spectrum is not
quantized. 
This implies that 
the discussions of the particle representation or the diagonalization of
the Hamiltonian are absent, while they are essential for the quantum
field theory.

There are two nonequilibrium extension of the thermal field theory,
i.e., the closed time path formalism \cite{CTP} and TFD
\cite{AIP}. The CTP formalism is widely used. But we employ the TFD
formalism in this paper, because  the concept of quasi-particle picture
 is clear even in nonequilibrium situations there.
In TFD, which is a real-time canonical formalism of quantum field
theory, thermal fluctuation is introduced through doubling the degrees
of the freedom, and the mixed state expectation is replaced by an
average of a pure state vacuum, called the thermal vacuum. 

It is crucial in our formulation of TFD to construct the interaction picture. 
In quantum field theory, the choice of unperturbed Hamiltonian and fields
 is that of quasi particle  picture, and concrete calculations are possible 
only when a particular unperturbed representation, or a particular
particle picture, is specified. One does not know an exact unperturbed
representation beforehand. Taking plausible representations, parameterized by
some parameters, we calculate the propagators of the Heisenberg fields
and require some conditions on them, called self-consistent renormalization
conditions, which pick up a self-consistent representation and determine the parameters.
The renormalized mass and coupling constant are such examples in quantum field theory.
 We construct a quasi particle picture in the doubled Fock space in TFD, defining quasi particle
operators which diagonalize the unperturbed TFD Hamiltonian.  In nonequilibrium 
case  a time-dependent number distribution is introduced as a unknown parameter, and 
a self-consistent renormalization condition 
derives an equation for it, i.e., the quantum Boltzmann equation \cite{AIP}. 
Moreover, the non-Markoffian extension of the self-consistent renormalization condition
is also proposed \cite{Chu3}. 
However, the extension and application to intrinsically inhomogeneous systems 
and condensed ones have not been established.

In this paper, we derive the non-Markoffian quantum transport
equations for cold atoms in a confining potential both without and with
a condensate from the nonequilibrium TFD formalism \cite{AIP,Chu3}.  
We confirm that our non-Markoffian transport equation for the non-condensed system is 
reduced to the ordinary quantum Boltzmann equation derived in the other methods 
when the Markov approximation is applied.
For the condensed system, we find that the non-Markoffian equation 
 contains an additional collision term which is overlooked in the other methods. 
This term vanishes in the equilibrium limit if there is no Landau instability, but 
 remains non-vanishing to prevent the system from equilibrating if there is Landau instability.
Thus our transport equation with the additional term (we call it the triple production term)
 and the other ones without it  
predict definitely different behaviors of the unstable system. This difference is traced back to
different quasi particle pictures in the respective theories.
Although we only consider in this paper the systems of trapped Bose atoms, 
our formulation of nonequilibrium TFD can be extended straightforwardly to the trapped 
systems of Fermi or multi-component atoms.

This paper is organized as follows. We briefly review the formulation of
nonequilibrium TFD in Sec.~II. In Sec.~III and IV, the non-condensed and
condensed systems are considered, respectively. We formulate each interaction
picture corresponding each quasi particle picture, diagonalizing 
the free (unperturbed) Hamiltonians of TFD.
The tensor form \cite{tensor}, which makes the diagrammatic calculation very
simple, is introduced for the non-condensed system, and is extended to
the condensed system. Applying the self-consistent renormalization
condition proposed by Chu and Umezawa \cite{Chu3}, we construct a
systematical method to obtain the transport equation for the trapped
systems. Section V is devoted to summary and discussions.


\section{Nonequilibrium TFD Formulation}
Here we briefly review the formulation of nonequilibrium TFD \cite{AIP}.

In TFD, every operator $A$ gets its tilde conjugation pair
$\Tilde{A}$\,, which is related to the ordinary (non-tilde) operator by
the following tilde conjugation rules: 
\begin{align}
	(AB)\Tilde{\phantom{i}} &= \Tilde{A} \Tilde{B} \,,\\
	(c_1 A + c_2B)\Tilde{\phantom{i}} &= c_1^* \Tilde{A} +
         c_2^*\Tilde{B} \,,\\ 
	(\Tilde{A})\Tilde{\phantom{i}} &= A \,,\\
	(A^\d)\Tilde{\phantom{i}} &= \Tilde{A}^\d \,,\\
	\ket\Tilde{\phantom{i}} &= \ket \,,\\
	\bra\Tilde{\phantom{i}} &= \bra \,,
\end{align}
where $c_1$ and $c_2$ are arbitrary c-numbers, and $\ket$ and $\bra$ are
the thermal vacua. The Hamiltonian of TFD, which should generate the
time translations of both non-tilde and tilde operators, is not the
ordinary Hamiltonian $H$ but the hat Hamiltonian $\Hat{H} = H -
\Tilde{H}$\,. The time independence of the thermal vacua requires the
minus sign in front of $\Tilde{H}$\,. 

The construction of the interaction picture is crucial, because the
choice of an interaction picture corresponds to that of a quasi particle
picture. Suppose the bosonic $a_\ell$-operators in the interaction
picture, representing the renormalized quasi particle with a quantum
number $\ell$.  They are related to the $\xi_\ell$-operators (called the
representation particle operators), which annihilate the time
independent thermal vacuum $\xi_\ell \ket  = 0$, through the thermal
Bogoliubov transformations 
\begin{alignat}{2}
	a^\mu_\ell(t)    &= B^{-1, \mu\nu}_\ell(t) \xi^\nu_\ell(t) \,,\\
	\ab^\nu_\ell(t)  &= \xib^\mu_\ell(t) B^{\mu\nu}_\ell(t) \,,\\
	\xi^\mu_\ell(t)  &= B^{\mu\nu}_\ell(t) a^\nu_\ell(t) \,,\\
	\xib^\nu_\ell(t) &= \ab^\mu_\ell(t) B^{-1, \mu\nu}_\ell(t) \,.
\end{alignat}
Here we introduce the thermal doublet notations
\begin{alignat}{2}
	a^\mu    &= \BM a \\ \at^\d \EM^\mu \,,& \hspace{0.5cm}
	\ab^\nu  &= \BM a^\d & -\at \EM^\nu \,,\\
	\xi^\mu  &= \BM \xi \\ \xit^\d \EM^\mu \,,&
	\xib^\nu &= \BM \xi^\d & -\xit \EM^\nu \,,
\end{alignat}
and the thermal Bogoliubov matrix
\begin{alignat}{2} 
	\label{ThermalBogoliubov1}
	B_\ell^{\mu\nu}(t)      &= \BM 1+ n_\ell(t) & - n_\ell(t) \\ -1
                                & 1 \EM^{\mu\nu} \,,\\ 
	\label{ThermalBogoliubov2}
	B^{-1, \mu\nu}_\ell(t) &= \BM 1 &  n_\ell(t) \\ 1 & 1+ n_\ell(t)
        \EM^{\mu\nu} \,. 
\end{alignat}
It is important to take the above particular form of the thermal
Bogoliubov matrix, as one calls $\alpha=1$ representation \cite{AIP},
which enables us to make use of the Feynman diagram method in
nonequilibrium systems \cite{Evans}. The number distribution $n_\ell(t)$
is given by 
\be
	n_\ell(t) = \bra a_\ell^\d(t) a_\ell(t) \ket \,,
\ee 
and its time dependence is determined later.

The unperturbed Hamiltonian for the $\xi_\ell$-operators should be
diagonal, consistently with the time independence of the thermal
vacuum. So the time dependence of $\xi_\ell$-operator in the interaction
picture should be in the form 
$
	\xi_\ell^\mu(t) = \xi_\ell^\mu e^{-i\int^t \!ds\;
	\omega_\ell(s)} \,, 
$
generated by the free Hamiltonian
$
	\Hat{H}_0(t) = \sum_\ell \omega_\ell (t) \xib_\ell^\mu(t)
	\xi_\ell^\mu(t) \, . 
$
Throughout this paper $\hbar$ is set to be unity. 
Note that $\omega_\ell$ generally depends on time because of the time
dependent energy renormalization. 
In this paper, we take time
independent $\omega_\ell$, assuming that the energy shift is negligible
in the leading order of perturbation, i.e., $ \xi_\ell^\mu(t) =
\xi_\ell^\mu e^{-i\omega_\ell t}$ and $\Hat{H}_0 = \sum_\ell \omega_\ell
\xib_\ell^\mu(t) \xi_\ell^\mu(t)$.
The unperturbed Hamiltonian for the $a_\ell$-operators is not
$\Hat{H}_0$,
but $\Hat{H}_Q(t) = \Hat{H}_0 - \Hat{Q}(t)$ with the thermal counter
term $\Hat{Q}(t)$ 
\begin{align}	
	\Hat{Q}(t)
	&=i \sum_\ell \Dot{n}_\ell(t) \ab_\ell^\mu(t) \BM 1 & -1 \\ 1 &
 -1 \EM^{\mu\nu}  a_\ell^\nu(t) \\\label{Q_a} 
	&= -i\sum_\ell \Dot{n}_\ell(t) \xib_\ell^\mu(t)\BM 0&1\\0&0
 \EM^{\mu\nu} \xi_\ell^\nu(t) \,, 
\end{align}
caused by the $t$-dependence of $n_\ell(t)$.

The field operator $\psi(x)$ is expanded with a complete set $\{
u_\ell(\bx)\}$ as 
\be \label{psiexpansion}
	\psi(x) = \sum_\ell a_\ell(t) u_\ell(\bx)\,, 
\ee
where $x = (\bx, t)$.
The unperturbed and full propagators for $\psi$ and $\xi$ are defined by
\begin{align}
	\Delta^{\mu\nu}(x,x') &= -i \bra {\rm T}[\psi^{\mu}(x) \,
 \Bar\psi^{\nu}(x')] \ket \,,\\ 
	G^{\mu\nu}(x,x') &= -i \bra {\rm T}[\psi^{\mu}_\MH(x) \,
 \Bar\psi^{\nu}_\MH(x')] \ket \,,\\ 
	d^{\mu\nu}_{\ell\ell'}(t,t') &= -i \bra {\rm
 T}[\xi^{\mu}_\ell(t) \, \xib^{\nu}_{\ell'}(t')] \ket \,,\\  
	g^{\mu\nu}_{\ell\ell'}(t,t') &= -i \bra {\rm
 T}[\xi^{\mu}_{\MH\ell}(t) \, \xib^{\nu}_{\MH\ell'}(t')] \ket \,, 
\end{align}
respectively, which are related to each other as
\begin{align} \label{eq:Delta}
	\Delta^{\mu\nu}(x,x') &= \sum_{\ell\ell'} 
	u_\ell(\bx) B_\ell^{-1, \mu\mu'}(t)  \notag\\[-8pt]
 &\hspace{1.5cm} \times  
	d^{\mu'\nu'}_{\ell\ell'}(t,t') B_{\ell'}^{\nu'\nu}(t')
 u_{\ell'}^*(\bx')\,,\\[8pt] 
	G^{\mu\nu}(x,x') &= \sum_{\ell\ell'} 
	u_\ell(\bx) B_\ell^{-1, \mu\mu'}(t)  \notag\\[-8pt]
 &\hspace{1.5cm} \times  
	g^{\mu'\nu'}_{\ell\ell'}(t,t') B_{\ell'}^{\nu'\nu}(t')
 u_{\ell'}^*(\bx')\,. 
\end{align}
The subscript ${}_\MH$ denotes a quantity in the Heisenberg picture.
While the unperturbed propagator $d$ has a diagonal structure
\be
	d^{\mu\nu}_{\ell\ell'}(t,t') = \delta_{\ell\ell'} \BM
	-i\theta(t-t') & 0 \\ 0 & i\theta(t'-t) \EM^{\mu\nu}
	e^{-i\omega_\ell(t-t')}\,, 
\ee
the full propagator $g$ has an upper triangular structure in general,
that is, $g^{12}_{\ell\ell'}(t,t') \ne 0$ and $g^{21}_{\ell\ell'}(t,t')
= 0$ \cite{AIP}. This is because $\xi_\MH^\d$ and $\xit_\MH^\d$
identically annihilate the bra-vacuum in the $\alpha=1$ representation
while $\xi_\MH$ and $\xit_\MH$ do not generally annihilate the
ket-vacuum. It is also shown that $g^{11}$ and $g^{22}$ are a retarded
and an advanced functions, respectively, as $d^{11}$ and $d^{22}$ are.

According to the Feynman method, one can calculate the full propagator
 with the interaction Hamiltonian in the interaction picture,
\be
{\hat H}_I ={\hat H}_{\rm int}+{\hat Q} \, ,
\ee
with ${\hat H}_{\rm int}={\hat H}-{\hat H}_0\,$. Possible
renormalization counter terms are suppressed below for simplicity.

\section{Trapped Bose Atoms in Non-Condensed System}

In this section, we derive the transport equation for the system of cold
Bose atoms without condensate
in nonequilibrium TFD. 

We start with the following Hamiltonian to describe the trapped dilute
Bose atoms, 
\be \label{Hamiltonian}
	H = \int\!\! d^3x\; 
	\left[ \psi^\d \left( -\frac{1}{2m}\nabla^2 + V(\bx) -\mu
	\right) \psi  + g\psi^\d\psi^\d\psi\psi\right] \,, 
\ee
where $m$, $V(\bx)$, $\mu$, and $g$ represent the mass of an atom, the
trap potential, the chemical potential, and the coupling constant,
respectively. The bosonic field operator $\psi(x)$ obeys the canonical
commutation relations 
\begin{align}
	[ \psi(x), \psi^\d(x')] \big|_{t=t'} &= \delta(\bx -\bx') \,,\\
	[ \psi(x), \psi(x')] \big|_{t=t'} &= [ \psi^\d(x), \psi^\d(x')]
 \big|_{t=t'}=0 \,. 
\end{align}
We expand the field
operator $\psi(x)$ as in Eq.~(\ref{psiexpansion}), using the solutions
of the following eigenequations, $\{ u_\ell(\bx)\}$ with the eigenvalues
$\{ \omega_\ell\}$, 
\be
	\left( -\frac{1}{2m}\nabla^2 + V(\bx) -\mu \right) u_\ell(\bx) =
	\omega_\ell u_\ell(\bx) \,.\ee  
The annihilation- and creation-operators $a_\ell$ and $a_\ell^\d$
diagonalize the free Hamiltonian part $H_0$ 
\be
	H_0 = \int\!\! d^3x\; \psi^\d \left( -\frac{1}{2m}\nabla^2 +
	V(\bx) -\mu \right) \psi  
	= \sum_\ell \omega_\ell a_\ell^\d a_\ell \,.
\ee

We apply the formulation of nonequilibrium TFD in the previous section
to the present system: Each degree of freedom is doubled, the time
dependent thermal Bogoliubov transformation is introduced in the
interaction picture, and the total Hamiltonian ${\hat H}$ is divided
into the unperturbed and interaction parts, ${\hat H}_Q$ and ${\hat
H}_I\, $. Then the full propagator is calculated in the Feynman diagram
method.

The self-consistent renormalization condition on the full propagator
thus obtained, which is extended from the self-consistent on-shell
renormalization condition in the ordinary quantum field theory, is
already proposed \cite{Chu1, Chu2, Chu3} as  
\be \label{RenormCond}
	g^{12}_{\ell\ell}(t,t) = 0 \, .
\ee
It provides the transport equation which determines the temporal
evolution of the unperturbed number distribution $n_\ell(t)$\,. 
Following the Dyson equations $G = \Delta + \Delta \Sigma G$ or $g = d +
d S g$, we obtain 
\be \label{Dyson_g12}
	g_{\ell\ell'}^{12}(t,t') = \sum_{mm'}\int\!\! ds ds' \; g_{\ell
	m}^{11}(t, s) S_{mm'}^{12}(s, s') g_{m'\ell'}^{22}(s', t') \,, 
\ee
with the self-energy
\begin{align}
	\Sigma^{\mu\nu}(x,x') &= \sum_{\ell\ell'} 
	u_\ell(\bx) B_\ell^{-1, \mu\mu'}(t) \notag\\[-8pt]
 &\hspace{1.5cm}\times  
	S^{\mu'\nu'}_{\ell\ell'}(t,t') B_{\ell'}^{\nu'\nu}(t')
 u_{\ell'}^*(\bx')\,,\\[8pt] 
	S^{\mu\nu}_{\ell\ell'}(t,t') &= 
	\BM S^{11}_{\ell\ell'}(t,t') & S^{12}_{\ell\ell'}(t,t') \\ 0 &
 S^{22}_{\ell\ell'}(t,t')\EM^{\mu\nu} \,. 
\end{align}

To illustrate how the transport equation follows from the
renormalization condition, we approximate the full propagators $g^{11
\,(22)}$ in Eq.~(\ref{Dyson_g12}) by the unperturbed ones $d^{11
\,(22)}$ and divide the self-energy $S$ into a loop contribution
$S_{\mathrm{loop}}$ and a contribution of the thermal counter term $S_Q$,
\be
	S_{Q\ell\ell'}^{\mu\nu}(t,t') =-i\Dot{n}_\ell(t)
	\delta_{\ell\ell'}\delta(t-t')\, \BM 0 & 1\\ 0 & 0 \EM^{\mu\nu}
	\,. 
\ee
Then we have
\begin{multline}
	g^{12}_{\ell\ell}(t, t) = -i \int_{-\infty}^t\!\!\!ds\; \Biggl[
 \Dot{n}_\ell(s) \\ 
	-2\Real \int_{-\infty}^{s} \!\!\!ds' \; e^{i\omega_\ell(s-s')}
 \, S_{\ell\ell, \mathrm{loop}}^{12}(s, s') \Biggr] \,, 
\end{multline}
and the renormalization condition (\ref{RenormCond}) implies the
following transport equation 
\be
	\Dot{n}_\ell (t) = 2\Real \int_{-\infty}^{t} \!\!\!ds \;
	e^{i\omega_\ell(t-s)} \, S_{\ell\ell, \mathrm{loop}}^{12}(t, s)
	\,. 
\ee

In view of this, we are going to calculate the self-energy perturbatively to obtain the transport equation in the leading order. 
Before that, we introduce the following tensor form \cite{tensor}
which makes the representations and calculations  of propagators and self-energies much 
more concise\, 
\be
	\begin{Bmatrix} s_1 \\ s_2 \end{Bmatrix}^{\mu} = 
	\left\{ \; \begin{matrix} s_1 & (\text{if}\quad \mu=1) \\ s_2 & (\text{if}\quad \mu = 2)\end{matrix} \right. \,.
\ee
For instance, the following matrix appearing in the unperturbed propagator is expressed in the tensor form as
\begin{align}
	\left[ B(t)^{-1} \BM 1&0\\0&0 \EM B(t') \right]^{\mu\nu} 
	&= \BM 1+n(t') & -n(t') \\ 1+n(t') & -n(t')\EM^{\mu\nu} \\
	&= \tf{1}{1}{1+n(t')}{-n(t')}\mu\nu \,.
\end{align}
Thus, the unperturbed propagators can be written in the tensor form as
\begin{align}
	&d^{\mu\nu}_{\ell\ell'}(t,t') = -i\delta_{\ell\ell'} e^{-i\omega_\ell(t'-t)} \notag\\ &\quad\times \biggl[
	\theta(t-t')\tf{1}{0}{1}{0}\mu\nu -\theta(t'-t) \tf{0}{1}{0}{1}\mu\nu \biggr]  \,, \\
	&\Delta^{\mu\nu}(x,x') = -i \ve^\nu \sum_\ell u_\ell(\bx) u_\ell^*(\bx') e^{-i\omega_\ell(t-t')} \notag\\
	&\quad\times \biggl[ 
	\theta(t-t') \tf{1}{1}{1+n_\ell(t')}{n_\ell(t')}\mu\nu \notag\\
	&\hspace{2cm} + \theta(t'-t) \tf{n_\ell(t)}{1+n_\ell(t)}{1}{1}\mu\nu
	\biggr]   \,,
\end{align}
with the sign factor, $ \ve^1 = 1$ and $\ve^2 = -1$\,.

As an example of manipulating products of the unperturbed propagators
 in the Feynman diagram calculation, we give the following manipulation,
\begin{align}
	&\phantom{=}
	\left[ B_{\ell }(t)^{-1} \BM 1&0\\0&0 \EM B_{\ell }(t') \right]^{\mu\nu} 
	\left[ B_{\ell'}(t)^{-1} \BM 1&0\\0&0 \EM B_{\ell'}(t') \right]^{\mu\nu} \notag\\
	&= \tf{1}{1}{1+n_{\ell}(t')}{-n_{\ell}(t')}\mu\nu \tf{1}{1}{1+n_{\ell'}(t')}{-n_{\ell'}(t')}\mu\nu \\
	&= \tf{1}{1}{(1+n_{\ell}(t'))(1+n_{\ell'}(t'))}{n_{\ell}(t')n_{\ell'}(t')}\mu\nu \,,
\end{align}
where the indices $\mu$ and $\nu$ are not summed over. This kind of manipulation
will be used below in the calculations of the self-energies.

\begin{figure}[t]
\includegraphics[width=5.5cm]{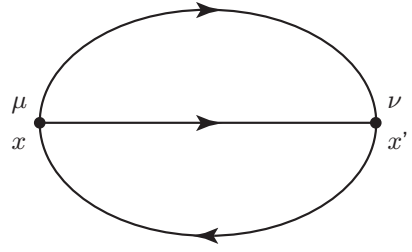}
\caption{\footnotesize{%
Two-loop self-energy diagram.
}}
\label{two-loop-Sigma}
\end{figure}

We focus on the two-loop self-energy indicated in
Fig.~\ref{two-loop-Sigma}, the leading loop diagram which makes
$g^{12}(t,t')$ nonzero, 
\be
	\Sigma_{\mathrm{loop}}^{\mu\nu}(x, x') = 
	- 2 g^2\ve^{\mu} \ve^{\nu} \Delta^{\mu\nu}(x,
	x')\Delta^{\mu\nu}(x, x')\Delta^{\nu\mu}(x', x) \,. 
\ee
The sign factors arise because of the definition of $\Bar\psi$ and of
the particular form of the interaction Hamiltonian 
$
 \Hat{H}_I = H_I - \Tilde{H}_I 
$\,.
Using the tensor form, we can rewrite the self-energy as
\begin{align}
	&\Sigma_{\mathrm{loop}}^{\mu\nu}(x, x') = -2i g^2 \ve^\nu
 \sum_{\ell_1\ell_2\ell_3}  
	e^{-i(\omega_{\ell_1}+\omega_{\ell_2}-\omega_{\ell_3})(t-t')}
 \notag\\ &\hspace{1.5cm}\times 
	u_{\ell_1}(\bx)u_{\ell_2}(\bx)u_{\ell_3}^*(\bx)
 u_{\ell_1}^*(\bx')u_{\ell_2}^*(\bx')u_{\ell_3}(\bx') \notag\\ 
	&\times \Biggl[
	\theta(t-t')
 \tf{1}{1}{(1+n_{\ell_1})(1+n_{\ell_2})n_{\ell_3}}{n_{\ell_1}n_{\ell_2}(1+n_{\ell_3})}\mu\nu 
 \notag\\ 
	&\hspace{0.4cm}+\theta(t'-t)
 \tf{n_{\ell_1}n_{\ell_2}(1+n_{\ell_3})}{(1+n_{\ell_1})(1+n_{\ell_2})n_{\ell_3}}{1}{1}\mu\nu  
	\Biggr]_{\min(t,t')}\,, 
\end{align}
where the subscript $\min(t,t')$ denotes the time arguments of $n$. So
that 
\begin{align}
	&S_{\mathrm{loop}, \ell\ell}^{12}(t,t') = 2ig^2
 \sum_{\ell_1\ell_2\ell_3}  
	e^{-i(\omega_{\ell_1} + \omega_{\ell_2} -
 \omega_{\ell_3}-\omega_{\ell})(t-t')} \notag\\ 
	&\hspace{0.3cm}\times C_{\ell_1\ell_2;\ell_3\ell} 
	\Bigl[n_{\ell_1}n_{\ell_2}(1+n_{\ell_3})(1+n_{\ell}) \notag\\  
	&\hspace{2.6cm}-
 (1+n_{\ell_1})(1+n_{\ell_2})n_{\ell_3}n_{\ell}\Bigr]_{\min(t,t')} \, , 
\end{align}
where
\be
	C_{\ell_1\ell_2;\ell_3\ell} = 
	\left| \int\!\! d^3x\; u_{\ell_1}(\bx) u_{\ell_2}(\bx)
	u_{\ell_3}^*(\bx) u_{\ell}^*(\bx) \right|^2 \,. 
\ee
Thus the transport equation at the two-loop level is derived
\begin{align} \label{transportEq}
	&\Dot{n}_\ell(t) 
	= 4g^2 \Real\!\! \int_{-\infty}^{t} \!\!\!ds \!\!
 \sum_{\ell_1\ell_2\ell_3} \!\! 
	e^{-i(\omega_{\ell_1} + \omega_{\ell_2} -
 \omega_{\ell_3}-\omega_\ell)(t-s)}  \,
 C_{\ell_1\ell_2;\ell_3\ell}\notag\\  
	&\times \Bigl[n_{\ell_1}n_{\ell_2}(1+n_{\ell_3})(1+n_{\ell}) -
 (1+n_{\ell_1})(1+n_{\ell_2})n_{\ell_3}n_{\ell}\Bigr]_s \,. 
\end{align}
The  transport equation of Markoffian type corresponding to Eq.~(\ref{transportEq}) has
already been derived by Chu and Umezawa \cite{Chu3} for the homogeneous
system. We have here extended their result to the inhomogeneous system
due to the confining potential. 

The exponential term $e^{-i(\omega_{\ell_1} + \omega_{\ell_2} -
\omega_{\ell_3}-\omega_\ell)(t-s)}$ in Eq.~(\ref{transportEq}) signifies
the energy conservation. To see that explicitly, we perform the time
integral in Eq.~(\ref{transportEq}) as 
\begin{align}
	\Dot{n}_\ell(t) &=
	4g^2 \Real\int_{-\infty}^{0} \!\!\!ds \;
 \sum_{\ell_1\ell_2\ell_3}  
	e^{i(\omega_{\ell_1} + \omega_{\ell_2} -
 \omega_{\ell_3}-\omega_{\ell} -i\frac{d}{dt})s} \notag\\ 
	&\hspace{3.5cm} \times C_{\ell_1\ell_2;\ell_3\ell}\;
 R_{\ell_1\ell_2;\ell_3\ell}(t) \\[10pt] 
	&= 4g^2  \sum_{\ell_1\ell_2\ell_3} 
	\frac{\Gamma_{\ell_1\ell_2;\ell_3\ell}(t)
 \;C_{\ell_1\ell_2;\ell_3\ell}\; R_{\ell_1\ell_2;\ell_3\ell}(t)}%
	{(\omega_{\ell_1} + \omega_{\ell_2} -
 \omega_{\ell_3}-\omega_{\ell})^2+\Gamma_{\ell_1\ell_2;\ell_3\ell}^2(t)}\,, 
\end{align}
where 
\begin{align}
	R_{\ell_1\ell_2;\ell_3\ell} &= 
	n_{\ell_1}n_{\ell_2}(1+n_{\ell_3})(1+n_{\ell}) \notag\\
 &\hspace{1.5cm}-
 (1+n_{\ell_1})(1+n_{\ell_2})n_{\ell_3}n_{\ell}\,,\\[5pt] 
	\Gamma_{\ell_1\ell_2;\ell_3\ell} &=
 -\frac{1}{R_{\ell_1\ell_2;\ell_3\ell}}
 \frac{dR_{\ell_1\ell_2;\ell_3\ell}}{dt}\,. 
\end{align}
It is seen from this Lorentzian form that the energy is conserved with
the width $\Gamma$ which is small when the temporal change of $n$ is
slow. 

To confirm the correspondence of our transport equation with those
derived in the other methods, we approximately replace $n_{\ell_i}(s)$
in the right hand side of Eq.~(\ref{transportEq}) with  $n_{\ell_i}(t)$,
in other words, take the limit $\Gamma\to0$ or the Markoffian limit,
with the assumption that the system is close to the equilibrium and the
evolution is sufficiently slow, 
\begin{align}
	&\Dot{n}_\ell(t) 
	= 4\pi g^2 \sum_{\ell_1\ell_2\ell_3} \delta(\omega_{\ell_1} +
 \omega_{\ell_2} - \omega_{\ell_3} - \omega_{\ell})\;  
	 C_{\ell_1\ell_2;\ell_3\ell}\notag\\ & \times  
	\Bigl[n_{\ell_1}n_{\ell_2}(1+n_{\ell_3})(1+n_{\ell}) -
 (1+n_{\ell_1})(1+n_{\ell_2})n_{\ell_3}n_{\ell}\Bigr]_t \,. 
\end{align}
The equation has a form of the ordinary quantum Boltzmann equation and
is consistent with the one which has been obtained from the on-shell
renormalization condition for the homogeneous system in nonequilibrium
TFD \cite{AIP}. However, the naive approximation is not valid for a
trapped system with the discrete energy spectrum $\omega_\ell$. The
energy conservation forced by the delta function is too  strict to
allow any energy exchange of particles and, consequently, any time
evolution. One way to avoid this difficulty is to apply the coarse
graining treatment \cite{Gardiner3}. For instance, in the case of the
harmonic trap potential, $\delta(\omega_{\ell_1} + \omega_{\ell_2} -
\omega_{\ell_3} - \omega_{\ell})$ was replaced by
$\delta_{\omega_{\ell_1} + \omega_{\ell_2},  \omega_{\ell_3} +
\omega_{\ell}}/\Omega$ with the trap frequency $\Omega$\,\cite{Holland}, 
\begin{align} \label{usualBoltzmann}
	&\Dot{n}_\ell(t) = \frac{4\pi g^2}{\Omega}
 \sum_{\ell_1\ell_2\ell_3}  
	\delta_{\omega_{\ell_1} + \omega_{\ell_2},  \omega_{\ell_3} +
 \omega_{\ell}} \; C_{\ell_1\ell_2;\ell_3\ell}\notag\\ 
	& \times \Bigl[n_{\ell_1}n_{\ell_2}(1+n_{\ell_3})(1+n_{\ell}) -
 (1+n_{\ell_1})(1+n_{\ell_2})n_{\ell_3}n_{\ell}\Bigr]_t \,. 
\end{align}
Note that such simple replacement is valid only for the harmonic trap
system whose energy-level spacing is uniform.

Another way is a semi-classical formulation using the phase-space
distribution function $n(\bx, \bm{p}, t)$. Then the particle energy is
no longer the discrete one but the local continuous one $\omega(\bx,
\bm{p}) = p^2/2m + V(\bx) -\mu$, for which the difficulty
mentioned above does not arise. Although the transport equations,
derived in this manner by several authors \cite{Gardiner2, Yamashita,
Zaremba}, are in good agreements with the experiments of evaporative
cooling and formation of condensate, they are classical and can not
describe quantum fluctuations fully, and the particle picture is not
explicit. In the next section we will show the case where the
instability of condensate exists and the quasi particle spectrum plays a 
crucial role, for which the semi-classical treatment is not valid.

\section{Trapped Bose Atoms in Condensed System}

In this section, we consider the situation in which 
a condensate
exists and derive the transport equation. The field operator $\psi(x)$
is divided into a classical part $\zeta(\bx)$ and a quantum part
$\phi(x)$, reflecting the existence of the condensate. In a fully
nonequilibrium situation, the order parameter $\zeta = \bra \psi \ket$
should be time-dependent, however we consider only a situation near the
equilibrium and assume the time-independent order parameter throughout
this paper. 

The Hamiltonian (\ref{Hamiltonian}) is written as
\be
	H = H_0 + H_{\rm int} \,,
\ee
where
\begin{align}
	H_0 &= \int\!\! d^3x\; 
		\biggl[ \phi^\d \left( -\frac{\nabla^2}{2m} + V(\bx) -
 \mu + 2g|\zeta(\bx)|^2 \right) \phi \label{BEC_H_0}\notag\\ 
		&\hspace{2cm}+ \frac{g}{2}\left( \zeta^{*2}(\bx) \phi^2
 + \zeta^2(\bx) \phi^{\d,2}\right) \biggr] \,,\\ 
	H_{\rm int} &= g \int\!\! d^3x\; 
		\biggl[ \zeta^*(\bx) \phi^\d \phi^2 + \zeta(\bx)
 \phi^{\d,2} \phi  
	+ \frac{1}{2}\phi^{\d,2} \phi^2
		\biggr] \,,
\end{align}
and the first order term of $\phi(x)$ vanishes since $\zeta(\bx)$ is
required to satisfy the following Gross-Pitaevskii equation \cite{GP} at
the tree level 
\be
	\left( -\frac{\nabla^2}{2m} + V(\bx) - \mu + g|\zeta(\bx)|^2
	\right)\zeta(\bx) = 0 \,. 
\ee

Next, we briefly review the Bogoliubov-de Gennes (BdG) method which
diagonalizes the free Hamiltonian $H_0$. The BdG equations are
simultaneous eigenvalue equations given by \cite{Bogoliubov, deGennes,
Fetter}
\be 
	T \by_\ell(\bx)  =  \omega_\ell \by_\ell(\bx) \,.
\ee
Here the doublet notation is introduced as
\begin{align}
	\by_\ell(\bx) &= \BM y_\ell^1(\bx) \\ y_\ell^2(\bx) \EM \,,\\
	T  &= \BM \L & \M \\ -\M^* & -\L \EM \,,
\end{align}
where 
\begin{align}
	\L &=  -\frac{\nabla^2}{2m} + V(\bx) - \mu + 2g|\zeta(\bx)|^2 \,,\\
	\M &= g\zeta^2(\bx) \,.
\end{align}
It is known that the BdG equations have the zero eigenvalue mode whose
treatment needs attention. We disregard the zero mode for simplicity
though it can be included consistently \cite{Lewenstein, Matsumoto},
because the scope of this paper is confined to the time-independent
order parameter and the dynamical effects of the zero mode is not
dominant then. In addition, the eigenvalues can be complex since the
operator $T$ is non-Hermitian. The condition for the emergence of
complex eigenvalues in the BdG equations has been studied both
numerically \cite{Pu, Wu, Kawaguchi} and analytically \cite{Skryabin,
Taylor, Nakamura}, and the quantum field theoretical formulation has
also been discussed \cite{Mine}. The emergence of complex eigenvalues
implies the dynamical instability of the system, and a drastic temporal
change of the order parameter occurs then, which is out of our present
formulation. In this paper, we consider only the case where no complex
eigenvalue emerges. 

Eigenfunctions belonging to the non-zero real eigenvalues can be
orthonormalized under the indefinite metric as
\begin{align} 
	\int\!\! d^3x\; \by_{\ell}^\d(\bx) \sigma_3 \by_{\ell'}(\bx) &=
 \delta_{\ell\ell'} \,,\label{orthonormal_1}\\ 
	\int\!\! d^3x\; \bz_{\ell}^\d(\bx) \sigma_3 \bz_{\ell'}(\bx) &=
 -\delta_{\ell\ell'} \,,\\ 
	\int\!\! d^3x\; \by_{\ell}^\d(\bx) \sigma_3 \bz_{\ell'}(\bx) &=
 0 \,,\label{orthonormal_3} 
\end{align}
with $i$-th Pauli matrix $\sigma_i$. The function $\bz_\ell$, defined by
$\bz_\ell = \sigma_1 \by_\ell^*$, is an eigenfunction belonging to
$-\omega_\ell$, if $\by_\ell$ is an eigenfunction belonging to
$\omega_\ell$. It is convenient to rewrite the orthonormal conditions
(\ref{orthonormal_1})--(\ref{orthonormal_3}) with the $2\times2$ matrix
form as 
\be	
	\int\!\! d^3x\; W_\ell(\bx) \; W_{\ell'}^{-1}(\bx) =
	\delta_{\ell\ell'} \,, 
\ee
where
\begin{align}
	W_\ell(\bx) &= \sigma_3 \BM \by_\ell^\d(\bx) \\
 \bz_{\ell}^\d(\bx) \EM \sigma_3 \,,\\ 
	W_\ell^{-1}(\bx) &= \BM \by_\ell(\bx) & \bz_\ell(\bx) \EM \,. 
\end{align}
The completeness condition, 
\be
	\sum_{\ell} \left[\by_\ell(\bx) \by_\ell^\d(\bx')-\bz_\ell(\bx)
	\bz_\ell^\d(\bx')\right] 
	=\sigma_3 \delta(\bx-\bx')\,,
\ee
can be expressed as
\be
	\sum_\ell W_\ell^{-1}(\bx) \; W_\ell(\bx') = \delta(\bx-\bx') \,,
\ee
and then the field operators are expanded in the doublet form as
\begin{align}
	\phi^\alpha(x) &= \sum_\ell W_\ell^{-1, \alpha\beta}(\bx)
 b_\ell^\beta(t) \,,\\ 
	\Bar\phi^\beta(x) &= \sum_\ell \bb_\ell^\alpha(t)
 W_\ell^{\alpha\beta}(\bx) \,, 
\end{align}
where
\begin{alignat}{2}
	\phi^\alpha &= \BM \phi \\ \phi^\d \EM^\alpha \,,&
 \hspace{0.5cm} 
	\Bar\phi^\alpha &= \BM \phi^\d & -\phi \EM^\alpha \,,\\
	b_\ell^\alpha &= \BM b_\ell \\ b^\d_\ell \EM^\alpha \,,&
 \hspace{0.5cm} 
	\bb_\ell^\alpha &= \BM b^\d_\ell & -b_\ell\EM^\alpha \,.
\end{alignat}
The operators $b_\ell$ satisfy the canonical commutation relation 
$[ b_\ell , b_{\ell'}^\d ] = \delta_{\ell\ell'}$, and diagonalizes the
free Hamiltonian (\ref{BEC_H_0}) 
\begin{align}
	H_0 &= \frac{1}{2} \int\!\! d^3x\; \Bar\phi^\alpha(x)
 \;T^{\alpha\beta}\; \phi^\beta(x) \\ 
		&= \sum_\ell \omega_\ell b_\ell^\d b_\ell \,.
\end{align}

The operators $b_\ell$ annihilate the Bose-Einstein condensed vacuum and
the operation of the creation operators $b_\ell$ on the vacuum
constructs the Fock space at zero temperature. Therefore, to treat this
system in nonequilibrium TFD, we double the degrees of freedom as
follows 
\begin{align}
	\BM b_\ell \\ \bt_\ell^\d \EM &= B_\ell^{-1} \BM \xi_\ell \\
 \xit_\ell^\d \EM  \,,\label{BEC_TB1}\\ 
	\BM b^\d_\ell & -\bt_\ell \EM &= \BM \xi^\d_\ell & -\xit_\ell
 \EM B_\ell \,,\label{BEC_TB2} 
\end{align}
where the thermal Bogoliubov matrix is defined in
Eqs.~(\ref{ThermalBogoliubov1}) and (\ref{ThermalBogoliubov2}) with the
quasi particle distribution $n_\ell(t) = \bra b_\ell^\d(t) b_\ell(t)
\ket$. Note that the operators who annihilate the thermal vacuum are not 
the $b$-operators but the $\xi$-operators. The combination of the two
transformations, $\xi$ into $b$ and $b$ into $\phi$, involves the
$4\times4$ transformations, 
\begin{align}
	\BM b_\ell \\ \bt_\ell^\d \\ b_\ell^\d \\ \bt_\ell \EM &= 
	\left(\begin{array}{cc|cc}
		1 & n_\ell & & \\ 1 & 1+n_\ell & & \\ \hline
		 & & 1+n_\ell & 1 \\ & & n_\ell & 1  
	\end{array}\right)	
	\BM \xi_\ell \\ \xit_\ell^\d \\ \xi_\ell^\d \\ \xit_\ell \EM
 \,,\\ 
	\BM \phi \\ \T\phi^\d \\ \phi^\d \\ \T\phi \EM &= \sum_\ell
	\left(\begin{array}{cc|cc}
		y_\ell^1 & & z_\ell^1 & \\ & y_\ell^1 & & z_\ell^1 \\ \hline
		y_\ell^2 & & z_\ell^2 & \\ & y_\ell^2 & & z_\ell^2 
	\end{array}\right)	
	\BM b_\ell \\ \bt_\ell^\d \\ b_\ell^\d \\ \bt_\ell \EM \,,
\end{align}
where the blank elements denote zero.
It is convenient to introduce the quartet notations for $b_\ell$ as follows
\begin{align}
	b_\ell^{\mu\alpha} 
	&= \BM b_\ell^\mu \\[8pt] [\sigma_1\bt_\ell]^\mu \EM^\alpha
	= \BM b_\ell \\ \bt_\ell^\d \\ b_\ell^\d \\ \bt_\ell
 \EM^{\mu\alpha} \,,\\ 
	\bb_\ell^{\nu\beta}
	&= \BM \bb_\ell^\nu & [\T\bb_\ell \sigma_1]^\nu \EM^\beta 
	= \BM b_\ell^\d & -\bt_\ell & -b_\ell & \bt_\ell^\d
 \EM^{\nu\beta} \,, 
\end{align}
and in similar fashions for $\xi_\ell$ and $\phi$. Then, the $4\times4$
transformations can be written simply 
\begin{align}
	b_\ell^{\mu\alpha} &= \mathcal{B}_\ell^{-1, \mu\alpha\nu\beta}
 \xi_\ell^{\nu\beta} \,,& 
	\bb_\ell^{\nu\beta} &= \xi_\ell^{\mu\alpha}
 \mathcal{B}_\ell^{\mu\alpha\nu\beta} \,,\\ 
	\phi^{\mu\alpha} &= \sum_\ell \mathcal{W}_\ell^{-1,
 \mu\alpha\nu\beta} b_\ell^{\nu\beta} \,,& 
	\Bar\phi^{\nu\beta} &= \sum_\ell \bb_\ell^{\mu\alpha}
 \mathcal{W}_\ell^{\mu\alpha\nu\beta} \,, 
\end{align}
with the $4\times4$ thermal Bogoliubov and BdG inverse matrices
\begin{align}
	\mathcal{B}_\ell^{-1, \mu\alpha\nu\beta} 
	&= \delta_{\alpha 1} \delta_{\beta 1} B_\ell^{-1, \mu\nu}
	+ \delta_{\alpha 2} \delta_{\beta 2}
	\left(\sigma_1 B_\ell^{-1}\sigma_1\right)^{\mu\nu} \!,\\
	\mathcal{B}_\ell^{\mu\alpha\nu\beta} &= \delta_{\alpha
 1}\delta_{\beta 1} B_\ell^{\mu\nu}  
	+ \delta_{\alpha 2}\delta_{\beta 2}\left(\sigma_1
 B_\ell\sigma_1\right)^{\mu\nu}  \,,\\ 
	\mathcal{W}_\ell^{-1, \mu\alpha\nu\beta} &= \delta_{\mu \nu}
 W_\ell^{-1,\alpha\beta} \,,\\ 
	\mathcal{W}_\ell^{\mu\alpha\nu\beta} &= \delta_{\mu
 \nu}W_\ell^{\alpha\beta}\,. 
\end{align}

The existence of the condensate brings no crucial alteration in defining
the thermal counter term, because the thermal Bogoliubov transformations
Eqs.~(\ref{BEC_TB1}) and (\ref{BEC_TB2}) remain unchanged. Thus, the
thermal counter term is the same as Eq.~(\ref{Q_a}) 
\be
	\Hat{Q} = -i\sum_\ell \Dot{n}_\ell(t) \xib_\ell^\mu(t)\BM
	0&1\\0&0 \EM^{\mu\nu} \xi_\ell^\nu(t) \,. 
\ee
Note however that the number distribution is that of the quasi
particles. With the quartet notation, $\Hat{Q}$ can be written as
\be
	\Hat{Q} = -\frac{i}{2}\sum_\ell \Dot{n}_\ell(t) \, 
	\xib_\ell^{\mu\alpha}(t) 	\left(\begin{array}{cc|cc}
		 0&1& & \\ 0&0 & & \\ \hline & & 0 & 0 \\ & &1 &0 
	\end{array}\right)^{\mu\alpha\nu\beta} \xi_\ell^{\nu\beta}(t) \,.
\ee
The unperturbed and full propagators are given by
\begin{align}
	\Delta^{\mu\alpha\nu\beta}(x, x') &= -i \bra
 T[\phi^{\mu\alpha}(x) \Bar\phi^{\nu\beta}(x')] \ket \,,\\ 
	G^{\mu\alpha\nu\beta}(x, x') &= -i \bra
 T[\phi_\MH^{\mu\alpha}(x) \Bar\phi_\MH^{\nu\beta}(x')] \ket \,,\\ 
	d^{\mu\alpha\nu\beta}_{\ell\ell'}(t,t') &= -i\bra
 T[\xi^{\mu\alpha}_{\ell}(t) \xib^{\nu\beta}_{\ell'}(t')] \ket \,,\\ 
	g^{ \mu\alpha\nu\beta}_{\ell\ell'}(t,t') &= -i\bra
 T[\xi^{\mu\alpha}_{\MH\ell}(t) \xib^{\nu\beta}_{\MH\ell'}(t')] \ket
 \,. 
\end{align}

Following the renormalization condition for the non-condensed system in
Eq.~(\ref{RenormCond}), which has successfully led to the transport
equation, we propose the renormalization condition for the condensed
system as 
\be \label{RenormCondCond}
	g^{1121}_{\ell\ell}(t,t) = 0 \, .
\ee

Let us consider the Dyson equation
\begin{multline}
	\BM g^{11} & g^{12} \\ g^{21} & g^{22} \EM = \BM d^{11} & 0 \\ 0
 & d^{22} \EM  \\ + 
	\BM d^{11} & 0 \\ 0 & d^{22} \EM \BM S^{11} & S^{12} \\ S^{21} &
 S^{22} \EM \BM g^{11} & g^{12} \\ g^{21} & g^{22} \EM \,, 
\end{multline}
where the superscripts denote the BdG indices, $\alpha$ and
$\beta$. Every matrix elements in the above equation are $2\times 2$
matrices with the thermal indices, $\mu$ and $\nu$, which are implicit
for conciseness of notation. Solving the Dyson equation for $g^{11}$,
we obtain 
\be \label{Dyson_g11}
	g^{11} = d^{11} + d^{11} \mathcal{S} g^{11} \,,
\ee
with
\be
	\mathcal{S}^{\mu\nu} = \Bigl[ S^{11} + S^{12} \bigl(1 - d^{22}
	S^{22}\bigr)^{-1} d^{22} S^{21} \Bigr]^{\mu\nu}\,. 
\ee
Obviously, $\mathcal{S}$ and $g^{11}$ in Eq.~(\ref{Dyson_g11}) are upper
triangular matrices, corresponding to the structures of $g^{\mu\nu}$ and
$S^{\mu\nu}$ for the non-condensed system. Then from
Eq.~(\ref{Dyson_g11}) follows 
\be	\label{g1211}
	g^{1121}_{\ell\ell'}\!(t,t') \!= \!\sum_{mm'} \! \int\!\! ds
	ds'\;  
	g^{1111}_{\ell m}(t,s) \mathcal{S}^{12}_{mm'}(s,s')
	g^{2121}_{m'\ell'}(s', t')\,. 
\ee
Similarly as in the non-condensed case Eq.~(\ref{Dyson_g12}), let us
consider the leading order. We approximate the full propagators
$g^{\mu1\mu 1}$ by the unperturbed ones $d^{\mu1\mu 1}$ and
$\mathcal{S}^{12}$ by $S^{1121}$ in the right hand side of
Eq.~(\ref{g1211}). Since the contribution of the thermal counter term to
the self-energy becomes 
\be
	S_{Q, \ell\ell'}^{\mu\alpha\nu\beta}(t,t')=
	-i\Dot{n}_\ell(t)\delta(t-t') \delta_{\ell\ell'} 
	\left(\begin{array}{cc|cc}
		 0&1& & \\ 0&0 & & \\ \hline & & 0 & 0 \\ & &1 &0 
	\end{array}\right)^{\mu\alpha\nu\beta} \!,
\ee
we obtain the following transport equation from Eq.~(\ref{RenormCondCond})
\be	\label{QBE_BEC_formal}
	\Dot{n}_\ell (t) = 2\Real \int_{-\infty}^{t} \!\!\!ds \;
	S_{\ell\ell, \mathrm{loop}}^{1121}(t, s) \,e^{i\omega_\ell(t-s)}
	\,. 
\ee

The tensor form introduced in the previous section makes the
perturbative calculation very simple and systematic. The unperturbed
propagators are written as 
\begin{widetext}
\begin{align}
&d^{\mu\alpha\nu\beta}_{\ell\ell}(t,t') = \notag\\
	&-i\theta(t-t') \Biggl[  
	 \tf{1}{0}{1}{0}\mu\nu \tf{1}{0}{1}{0}\alpha\beta
 e^{-i\omega_\ell(t-t')} 
	 +\tf{0}{1}{0}{1}\mu\nu \tf{0}{1}{0}{1}\alpha\beta
 e^{i\omega_\ell(t-t')} \Biggr] \notag\\ 
	&+i\theta(t'-t) \Biggl[ 
	 \tf{0}{1}{0}{1}\mu\nu \tf{1}{0}{1}{0}\alpha\beta
 e^{-i\omega_\ell(t-t')} 
	+\tf{1}{0}{1}{0}\mu\nu \tf{0}{1}{0}{1}\alpha\beta
 e^{i\omega_\ell(t-t')} \Biggr] \,, 
\end{align}
\begin{align}
	\Delta^{\mu\alpha\nu\beta}(x,x') = -i \ve^\nu \ve^\beta
 \sum_\ell u_\ell(\bx) &u_\ell^*(\bx') \notag\\  
	\times\left(\vbox to 23pt{}\right.
	\theta(t-t') \Biggl[ 
	 &\tf{1}{1}{1+n_\ell(t')}{n_\ell(t')}\mu\nu 
	\biggl\{\by_\ell(\bx)\biggr\}^\alpha
 \biggl\{\by_\ell^*(\bx')\biggr\}^\beta  
	e^{-i\omega_\ell(t-t')} \notag\\
	+&\tf{1}{1}{n_\ell(t')}{1+n_\ell(t')}\mu\nu
	\biggl\{\bz_\ell(\bx)\biggr\}^\alpha
 \biggl\{\bz_\ell^*(\bx')\biggr\}^\beta  
	e^{i\omega_\ell(t-t')} \Biggr]_{t'}\notag\\[10pt]
	+\theta(t'-t)\Biggl[ 
	 &\tf{n_\ell(t)}{1+n_\ell(t)}{1}{1}\mu\nu 
	\biggl\{\by_\ell(\bx)\biggr\}^\alpha
 \biggl\{\by_\ell^*(\bx')\biggr\}^\beta  
	e^{-i\omega_\ell(t-t')} \notag\\
	+&\tf{1+n_\ell(t)}{n_\ell(t)}{1}{1}\mu\nu 
	\biggl\{\bz_\ell(\bx)\biggr\}^\alpha
 \biggl\{\bz_\ell^*(\bx')\biggr\}^\beta  
	e^{i\omega_\ell(t-t')} \Biggr]_t
	\left. \vbox to 23pt{} \right) \,.
\end{align}
The one-loop self-energy, indicated in Fig.~\ref{one-loop-Sigma} (a), is
 given in the tensor form as 
\begin{align}
	\Sigma_{\mathrm{loop}}^{\mu\alpha\nu\beta}&(x,x') = -2ig^2
 \ve^\nu \ve^\alpha \sum_{\ell_1\ell_2}  
	\left[\vbox to 23pt{}\right. \notag\\
	 &\theta(t-t')
 \tf{1}{1}{(1+n_{\ell_1})(1+n_{\ell_2})}{n_{\ell_1}n_{\ell_2}}\mu\nu 
	\biggl\{\chi_{yy}(\bx)\biggr\}^\alpha
 \biggl\{\chi_{yy}^*(\bx')\biggr\}^\beta  
	e^{-i(\omega_{\ell_1} + \omega_{\ell_2})(t-t')} \notag\\
	+&\theta(t'-t)\tf{n_{\ell_1}n_{\ell_2}}{(1+n_{\ell_1})(1+n_{\ell_2})}{1}{1}\mu\nu 
	\biggl\{\chi_{yy}(\bx)\biggr\}^\alpha \biggl\{\chi_{yy}^*(\bx')\biggr\}^\beta 
	e^{-i(\omega_{\ell_1} + \omega_{\ell_2})(t-t')} \notag\\
	+&\theta(t-t')
 \tf{1}{1}{(1+n_{\ell_1})n_{\ell_2}}{n_{\ell_1}(1+n_{\ell_2})}\mu\nu 
	\biggl\{\chi_{yz}(\bx)\biggr\}^\alpha
 \biggl\{\chi_{yz}^*(\bx')\biggr\}^\beta  
	e^{-i(\omega_{\ell_1} - \omega_{\ell_2})(t-t')} \notag\\
	+&\theta(t'-t)\tf{n_{\ell_1}(1+n_{\ell_2})}{(1+n_{\ell_1})n_{\ell_2}}{1}{1}\mu\nu 
	\biggl\{\chi_{yz}(\bx)\biggr\}^\alpha
 \biggl\{\chi_{yz}^*(\bx')\biggr\}^\beta  
	e^{-i(\omega_{\ell_1} - \omega_{\ell_2})(t-t')} \notag\\
	+&\theta(t-t')
 \tf{1}{1}{n_{\ell_1}(1+n_{\ell_2})}{(1+n_{\ell_1})n_{\ell_2}}\mu\nu  
	\biggl\{\chi_{zy}(\bx)\biggr\}^\alpha \biggl\{\chi_{zy}^*(\bx')\biggr\}^\beta 
	e^{i(\omega_{\ell_1} - \omega_{\ell_2})(t-t')} \notag\\
	+&\theta(t'-t)\tf{(1+n_{\ell_1})n_{\ell_2}}{n_{\ell_1}(1+n_{\ell_2})}{1}{1}\mu\nu 
	\biggl\{\chi_{zy}(\bx)\biggr\}^\alpha
 \biggl\{\chi_{zy}^*(\bx')\biggr\}^\beta  
	e^{i(\omega_{\ell_1} - \omega_{\ell_2})(t-t')} \notag\\
	+&\theta(t-t')
 \tf{1}{1}{n_{\ell_1}n_{\ell_2}}{(1+n_{\ell_1})(1+n_{\ell_2})}\mu\nu  
	\biggl\{\chi_{zz}(\bx)\biggr\}^\alpha
 \biggl\{\chi_{zz}^*(\bx')\biggr\}^\beta  
	e^{i(\omega_{\ell_1} + \omega_{\ell_2})(t-t')} \notag\\
	+&\theta(t'-t)\tf{(1+n_{\ell_1})(1+n_{\ell_2})}{n_{\ell_1}n_{\ell_2}}{1}{1}\mu\nu 
	\biggl\{\chi_{zz}(\bx)\biggr\}^\alpha
 \biggl\{\chi_{zz}^*(\bx')\biggr\}^\beta  
	e^{i(\omega_{\ell_1} + \omega_{\ell_2})(t-t')} 
	\left. \vbox to 23pt{} \right]_{\min(t, t')}\,,
\end{align}
\end{widetext}
where 
\begin{align}
	\chi^\alpha_{yy} &= 
	  \zeta^{\Bar\alpha}\by_{\ell_1}^\alpha \by_{\ell_2}^\alpha
	+ \zeta^{\alpha}\by_{\ell_1}^{\Bar\alpha} \by_{\ell_2}^\alpha
	+ \zeta^{\alpha}\by_{\ell_1}^\alpha \by_{\ell_2}^{\Bar\alpha}
	\,,\\
	\chi^\alpha_{yz} &= 
	  \zeta^{\Bar\alpha}\by_{\ell_1}^\alpha \bz_{\ell_2}^\alpha
	+ \zeta^{\alpha}\by_{\ell_1}^{\Bar\alpha} \bz_{\ell_2}^\alpha
	+ \zeta^{\alpha}\by_{\ell_1}^\alpha \bz_{\ell_2}^{\Bar\alpha}
	\,,\\
	\chi^\alpha_{zy} &= 
	  \zeta^{\Bar\alpha}\bz_{\ell_1}^\alpha \by_{\ell_2}^\alpha
	+ \zeta^{\alpha}\bz_{\ell_1}^{\Bar\alpha} \by_{\ell_2}^\alpha
	+ \zeta^{\alpha}\bz_{\ell_1}^\alpha \by_{\ell_2}^{\Bar\alpha}
	\,,\\
	\chi^\alpha_{zz} &= 
	  \zeta^{\Bar\alpha}\bz_{\ell_1}^\alpha \bz_{\ell_2}^\alpha
	+ \zeta^{\alpha}\bz_{\ell_1}^{\Bar\alpha} \bz_{\ell_2}^\alpha
	+ \zeta^{\alpha}\bz_{\ell_1}^\alpha \bz_{\ell_2}^{\Bar\alpha}
	\,,
\end{align}
with $\zeta^\alpha(\bx)=\BM \zeta(\bx) \\ \zeta^*(\bx) \EM^{\alpha}$, 
and $\Bar\alpha$ denotes $\Bar\alpha = 2, 1$ for $\alpha = 1, 2$,
respectively. Since
\begin{multline}
	S^{1121}_{\ell\ell}(t,t') =\int\!\!d^3xd^3x'\; 
	B_\ell^{1\mu}(t) W_\ell^{1\alpha}(\bx)  \\ 
	\times \Sigma_{\mathrm{loop}}^{\mu\alpha\nu\beta}(x,x')
	W_\ell^{-1, \beta 1}(\bx') B_\ell^{-1, \nu 2}(t) \,,
\end{multline}
we obtain from Eq.~(\ref{QBE_BEC_formal})
\begin{widetext}
\begin{figure*}[th]
\includegraphics[width=10.4cm]{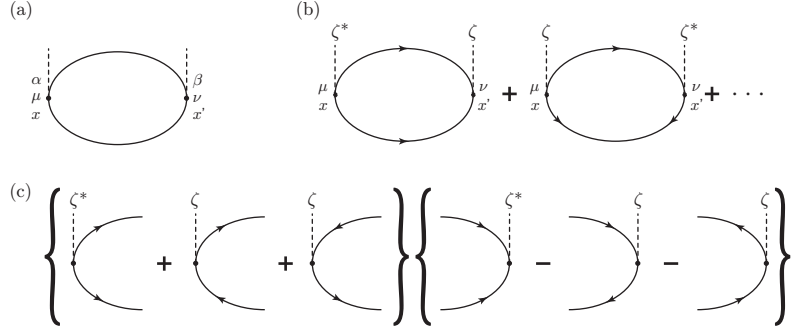}
\caption{\footnotesize{%
(a) One-loop self-energy diagram.\;
(b) A part of the one-loop self-energy diagrams for $\alpha = \beta
 =1$. There are nine terms in total.\; 
(c) A description in the tensor form for $\alpha = \beta =1$. The
 diagrams can be factorize in terms of tensor form. 
}}
\label{one-loop-Sigma}
\end{figure*}
\begin{align}	\label{BEC_QBE}
	\Dot{n}_\ell(t) = 4g^2\sum_{\ell_1, \ell_2} &\int_{-\infty}^{t}
 \!dt'\; \mathrm{Re} \Bigl[ \notag\\ 
	 &e^{i(\omega_\ell - \omega_{\ell_1} - \omega_{\ell2})(t-t')}
	|(\by_\ell, \chi_{yy})|^2 \left\{ (1+n_\ell)n_{\ell_1}n_{\ell_2}
 - n_\ell(1+n_{\ell_1})(1+n_{\ell_2})\right\} \notag\\ 
	+&e^{i(\omega_\ell - \omega_{\ell_1} + \omega_{\ell2})(t-t')}
	|(\by_\ell, \chi_{yz})|^2 \left\{ (1+n_\ell)n_{\ell_1}(1+n_{\ell_2}) - n_\ell(1+n_{\ell_1})n_{\ell_2}\right\} \notag\\
	+&e^{i(\omega_\ell + \omega_{\ell_1} - \omega_{\ell2})(t-t')}
	|(\by_\ell, \chi_{zy})|^2 \left\{
 (1+n_\ell)(1+n_{\ell_1})n_{\ell_2} - n_\ell
 n_{\ell_1}(1+n_{\ell_2})\right\} \notag\\[-2pt] 
	+&e^{i(\omega_\ell + \omega_{\ell_1} + \omega_{\ell2})(t-t')}
	|(\by_\ell, \chi_{zz})|^2 \left\{
 (1+n_\ell)(1+n_{\ell_1})(1+n_{\ell_2}) - n_\ell
 n_{\ell_1}n_{\ell_2}\right\} 
	\;\Bigr]_{t'} \,,
\end{align}
\end{widetext}
with
\be
	(\by_\ell, \chi) = \int\!\! d^3x\; \by_\ell^{*, \alpha}(\bx) \chi^{\alpha}(\bx) \,.
\ee
This is the non-Markoffian transport equation for the condensed system at one-loop level.
The first term in Eq.~(\ref{BEC_QBE}) corresponds to the Beliaev damping and its inverse process, 
and the second and third terms do to the Landau damping and their inverse processes.
These processes bring the system to the equilibrium.
On the other hand, the fourth term corresponds to a process
in which three quasi particles are created or annihilated.
We call it the triple production process.
The existence of the process prevents the system from equilibrating, 
because the collision term is always nonzero $ (1+n_\ell)(1+n_{\ell_1})(1+n_{\ell_2}) - n_\ell n_{\ell_1}n_{\ell_2} > 0$\,.
Note however that if all the energies of quasi particles are positive, 
the process is forbidden due to the energy conservation.
Conversely, once a negative energy mode exists, 
the excitations to the mode will proceed until the condensate decays.
This exactly corresponds to the scenario of the Landau instability in terms of the kinetics. 
Thus the triple production term is interpreted to induce the decay processes
 leading to the Landau instability. 

To see the importance of quantum field theoretical treatment based on the proper
quasi particle representation, 
let us treat the same system in the particle picture of original atoms.
Namely, we suppose that a macroscopic number of atoms is in the 
particular state $\ell = \ell_c$, and consider it in the transport equation Eq.~(\ref{transportEq})
derived for the non-condensed system.
The energy $\omega_\ell$ is shifted in such a manner that $\omega_{\ell_c}$ is vanishing, 
and the particle distribution $n_\ell(t)$ is replaced by $n_\ell(t) + \delta_{\ell_c \ell} N_c$\,.
If it is assumed that $N_c\simeq N$ with the total atom number $N$, the transport equation approximately becomes
\begin{widetext}
\begin{align}\label{NONBEC_QBE}
	\Dot{n}_\ell(t) \simeq 4g^2 N_c \sum_{\ell_1, \ell_2} &\int_{-\infty}^{t} \!dt'\; \mathrm{Re} \Bigl[ \notag\\
	 &e^{i(\omega_\ell - \omega_{\ell_1} - \omega_{\ell2})(t-t')}
	C_{\ell_1\ell_2;\ell_c\ell} \left\{ (1+n_\ell)n_{\ell_1}n_{\ell_2} - n_\ell(1+n_{\ell_1})(1+n_{\ell_2})\right\} \notag\\
	+&e^{i(\omega_\ell - \omega_{\ell_1} + \omega_{\ell2})(t-t')}
	C_{\ell_1\ell_c;\ell_2\ell} \left\{ (1+n_\ell)n_{\ell_1}(1+n_{\ell_2}) - n_\ell(1+n_{\ell_1})n_{\ell_2}\right\} \notag\\[-2pt]
	+&e^{i(\omega_\ell + \omega_{\ell_1} - \omega_{\ell2})(t-t')}
	C_{\ell_c\ell_2;\ell_1\ell} \left\{ (1+n_\ell)(1+n_{\ell_1})n_{\ell_2} - n_\ell n_{\ell_1}(1+n_{\ell_2})\right\}
	\;\Bigr]_{t'} \,.
\end{align}
\end{widetext}
This equation corresponds to Eq.~(\ref{BEC_QBE}), but the triple production term is absent.
Its absence comes from the inadequate choice of particle picture.

The difference between Eqs.~(\ref{BEC_QBE}) and (\ref{NONBEC_QBE}) is decisive
 when the system has the Landau instability.  We conclude that the Landau instability should
be described by Eq.~(\ref{BEC_QBE}) with the triple production
 term, because it is based on the appropriate
quasi particle picture. The omission of the triple production term, when the energy conservation
allows it, would violate the unitarity in quantum theory.

\section{Summary and Discussions}

In this paper, nonequilibrium TFD is applied to the systems of trapped Bose atoms, 
and the quantum transport equations of non-Markoffian type have been derived  for 
both the non-condensed and condensed systems.
We have diagonalized the unperturbed Hamiltonians, each of which corresponds 
to the quasi particle picture, and this diagonalization procedure in the interaction
picture is essential for TFD as well as for ordinary quantum field theory.
To derive the transport equations for the trapped systems both without and with a condensate, 
we have applied the self-consistent renormalization condition proposed by Chu and Umezawa 
for a homogeneous system. 
In order to make complicated calculations of the self-energies transparent, 
we have also refined the diagrammatic calculations in the tensor form, and have developed
the convenient $4 \times 4$-matrix formulation in the condensed case. 
Although the transport equations are derived only in the lowest order in this paper,
their higher order corrections  can be obtained systematically in our method,
simply by calculating higher order diagrams. In contrast, higher order corrections 
of the transport equations cannot be obtained straightforwardly in the other methods.

For the non-condensed system, the non-Markoffian transport equation at  two-loop level
 derived in this paper 
becomes very similar in the Markoffian limit to those derived in the different methods.
While the  equations in the other methods involve a delta function 
and require a strict energy conservation in each collision,
the energy in our equation is conserved with the finite width which reflects thermal changes
and can be calculated. With the strict conservation, the collision integral is either zero or infinite because of the delta function 
in a trapped system where the energy spectrum is discrete.
Therefore, an additional cure was needed to avoid the problem in the other methods.
Although the problem does not occur in the semi-classical method since
the delta function is integrated over continuous energy variable,
the semi-classical treatments are not consistent with the particle picture in the trapped system. 
It is remarked that our equation in nonequilibrium TFD follows from the correct particle picture
 and needs no additional patch.

Perturbative calculations  are much more intricate for the condensed system
 than for the non-condensed one: In the former the two component eigenfunctions
of the BdG equations complicate expressions.  We have merged the thermal doublet and
the BdG one into a quartet and have constructed the $4 \times 4$-matrix formalism,
which is helpful for our concrete calculations of nonequilibrium TFD.

In principle similarly as in the non-condensed case, we have derived the quantum
transport equation in the condensed case at  one-loop level. A crucial point in our
equation is that it involves an additional collision term, absent in the transport
equations of the other methods. Usual collision terms in the lowest order are only
those corresponding to the Beliaev and Landau damping and their inverse processes.
Our additional term represents creation or annihilation of three quasi particles,
and prevents the system from equilibrating if a negative energy mode exists 
and is suppressed otherwise. So the behavior of the system with the Landau instability,
described by our equation, is distinguished from those under the equations without the triple 
production term. We emphasize that the additional term disappears in the inadequate
 particle picture, as was shown at the end of Sec.~IV, but that it appears naturally in quantum
field theory.

As for the Landau instability, the authors of Ref.~\cite{Iigaya} have pointed out that 
the sign of the Landau damping rate changes to minus when the system has Landau instability, 
which is interpreted as an indication of the decay of the condensate.
In their analysis, the nonequilibrium distribution function was roughly approximated
 to the Bose-Einstein one.
However, this approximation is invalid for the negative energy spectrum, 
because the distribution function becomes negative and therefore unphysical.
In our scenario with the transport equation, the dominant term in the Landau instability 
is obviously the triple production one which is always positive.

We comment on a relation between the closed time path formalism (CTP) \cite{CTP}
 and our nonequilibrium TFD.  Both derive the very similar Dyson-Schwinger equations,
 indeed the same in form. This is because 
they follow from the common Heisenberg equations and Feynman diagrams. 
It is important to notice that the the same Dyson-Schwinger equation
does not always give the same solution. A main difference between the two
approaches is in their unperturbed propagators. The unperturbed propagator
in CTP is evaluated over the density matrix at the initial time $t_0$, out of equilibrium,
and would become $\displaystyle \Delta_{\rm CTP}^{\mu\nu}(x,x') \sim
 B^{-1, \mu\mu'}[n_\ell(t_0)] d^{\mu'\nu'}_{\ell\ell'}(t,t')
 B^{\nu'\nu}[n_{\ell'}(t_0)] $ instead of Eq.~(\ref{eq:Delta}) in TFD. 
Thus, while the thermal Bogoliubov matrix in the unperturbed propagator of CTP is time-independent
and carries information only on the initial thermal state, our propagator with the
time-dependent Bogoliubov matrix adopts temporal thermal changes of the system.
So the unperturbed representation in TFD contains some non-perturbative effects in perturbative
calculations of CTP. The time-dependence of the thermal Bogoliubov matrix in TFD
is important also in respect of the self-consistent renormalization, lacking in CTP so far:
The time-dependent thermal Bogoliubov matrix 
creates the thermal counter term $\hat Q$ in the interaction Hamiltonian, and 
the derivation of transport equations would be impossible without it.

The transport equation derived in this paper can describe only the initial stage of the 
condensate decay with the Landau instability.
It is out of our present formulation to describe a full time evolution of the decay 
as well as that of quantum phase transition through evaporative cooling,
because we have ignored the time dependence of the condensate involving the time dependent
quasi particle representation.
In such cases, the zero modes of the BdG equations play a crucial role and must be considered.
The description of the decay with the Landau instability, that with the dynamical instability
a full description of quantum phase transition are challenging subjects in nonequilibrium TFD.

\begin{acknowledgments}
This work is partly supported by 
Waseda University Grant for Special Research Projects
 (Project number: 2009A-886 and 2009B-147). 
The authors thank the Yukawa Institute for Theoretical Physics at Kyoto University 
for offering us the opportunity to discuss this work during the YITP workshop 
YITP-W-08-09 on ``Thermal Quantum Field Theories and Their Applications''.
\end{acknowledgments}


\end{document}